\documentclass[12pt,letterpaper]{article}
\usepackage{amsmath,amssymb,pgf,pgfarrows,pgfnodes,float,appendix, hyperref}
\usepackage{graphicx}
\usepackage{subfigure}
\usepackage[margin=0.9in]{geometry}

\def\slash#1{\ensuremath{\;/\!\!\!\! #1}}

\newcommand{\be}{\begin{equation}}
\newcommand{\ee}{\end{equation}}
\newcommand{\bea}{\begin{eqnarray}}
\newcommand{\eea}{\end{eqnarray}}

\title{{\rm\footnotesize \qquad \qquad \qquad \qquad \qquad \ \qquad \qquad \qquad \ \ \ \ \ \                       RUNHETC-2014-2     
SCIPP 14/01}\vskip.5in     Cosmological SUSY Breaking and the Pyramid Scheme}
\author{Tom Banks\\
Department of Physics and SCIPP\\
University of California, Santa Cruz, CA 95064\\
{\it and}\\
Department of Physics and NHETC\\
Rutgers University, Piscataway, NJ 08854\\
E-mail: \href{mailto:banks@scipp.ucsc.edu}{banks@scipp.ucsc.edu}
\\
}
\date{}
\begin{document}
\maketitle

\begin{abstract}
I review the ideas of holographic space-time (HST), Cosmological SUSY breaking (CSB), and the Pyramid Schemes, which are the only known models of Tera-scale physics consistent with CSB, current particle data, and gauge coupling unification. There is considerable uncertainty in the estimate of the masses of supersymmetric partners of the standard model particles, but the model predicts that the gluino is probably out of reach of the LHC, squarks may be in reach, and the NLSP is a right handed slepton, which should be discovered soon. \end{abstract}

\section{Introduction}

All known consistent string theory models, in asymptotically flat space-time, are exactly supersymmetric.  All well established examples of the $AdS/QFT$ correspondence, with radius that can be parametrically large in string units, give flat space limits that are exactly super-symmetric.

The world around us is not supersymmetric.   If a theory of quantum gravity is to have relevance to the world, it must explain to us how SUSY is broken, and extant string theory models do not do this.  The rest of the papers in this volume treat SUSY breaking in ``string theory" by taking a low energy quantum field theory derived as an approximation to a supersymmetric string model, and adding supersymmetry breaking terms to it, which are {\it at best} plausibly connected to excitations in string theory.  I have criticized this procedure extensively\cite{tbgadfly}.  I believe instead that SUSY breaking is connected to the asymptotic structure of space-time via degrees of freedom that are thrown away in effective field theory.  These are not high energy DOF, but rather very low energy excitations, which decouple from particles localized in the bulk because they are localized on the causal horizon.  In flat and AdS space, the horizon actually recedes to infinity, and the horizon DOF need not be included in the Hilbert space, but in de Sitter (dS) space, the finite horizon leads to a finite amount of SUSY breaking.

The theory of Holographic Space-time explains the ``empirical facts" of supersymmetry in string theory by choosing (in Minkowski space) the variables of quantum gravity to be the cut-off generators of the (generalized) super-BMS algebra.   The super-BMS algebras are defined on the lightcone in $4$ dimensional\footnote{We stick to four dimensions for simplicity of exposition.  The formalism generalizes to higher dimensions.} Lorentzian momentum space. It has two components, corresponding to the top and bottom of the cone.  The $P_0 > 0$ component has generators $\psi_{\alpha }^+ (P, a)$, satisfying

$$[ \psi_{\alpha }^+ (P, a), \bar{\psi}_{\dot{\beta} }^+ (Q, b)]_+ = Z_{ab} M_{\mu} (P,Q) \sigma^{\mu}_{\alpha\dot{\beta}} \delta (P \cdot Q) ,$$ and $$ \psi_{\alpha} (P, a) \sigma^{\mu}_{\alpha\dot{\beta}} P_{\mu}  = .0 $$

The delta function is non-zero only when $P$ and $Q$ are collinear, and so both positive multiples of $(1, {\bf \Omega})$, where ${\bf \Omega}$ is a point on the unit 2-sphere.  $M^{\mu}$ points in the same direction and chooses the minimum of the two positive multipliers.  The constraint on $\psi $ says that it lies in the holomorphic spinor bundle over the two sphere, while it's conjugate lies in the anti-holomorphic bundle.   The local super-algebra, at fixed $P$, completed by the commutators of $Z_{ab}$ with $psi (P, a)$, and with themselves, has a finite dimensional unitary representation generated by the action of the fermionic generators on a single state.  One should think of the index $a$ as labeling the eigen-functions of the Dirac operator on some compact manifold, with a cutoff that fixes the volume of that manifold in Planck units\cite{tbjk}.  Other geometric properties of the compact space are encoded in the super-algebra.  For the purposes of this note, we can just set the $Z_{ab}$ to be a c-number times $\delta_{ab} $ and let $a,b$ run over the 16 components of a spinor in $7$ dimensions.
This is the quantum theory of eleven dimensional supergravity compactified on a $7$ torus of Planck size.
For fixed $P$, the super-BMS algebra is just the algebra of a single eleven dimensional graviton, and its superpartners, compactified on a Planck sized torus, and throwing away states whose mass is larger than the Planck mass.  This includes all states charged under the multiple $U(1)$ gauge symmetries of ${\cal N} = 8$ SUGRA.

Scattering theory, at least in theories of gravity, is considered to be a map between the past and future representations of the super-BMS algebra
$$ \psi_{\alpha}^{(+)} = S^{\dagger} \psi_{\alpha}^{(-)} S\footnote{For the BMS sub-algebra, Strominger has interpreted (the bosonic square of) this equation as an expression of spontaneous breakdown of the BMS symmetry, and shown that Weinberg's graviton low energy theorems follow from this equation.}. $$   Scattering states are states in which the operator valued measures $\psi_{\alpha} (P,a)$ vanish outside of the endcaps of a finite number of Sterman-Weinberg cones, but are also non-vanishing at $P = 0$, except in small annuli surrounding the end-caps.  

For finite causal diamonds, these singular measures are replaced by a sum over a finite set of spinor spherical harmonics.  The $P = 0$ modes are the degrees of freedom responsible for the entropy of horizons.   For finite horizons they contribute finite terms to the Hamiltonian, but they decouple in the infinite horizon limit.  

Cosmological SUSY breaking (CSB) is an attempt to implement the consequences of these abstract ideas in low energy effective field theory, and use them to guess at the correct model of Tera-scale physics.  It leads to a quite restrictive set of models.  The phenomenological analysis of these models is difficult because they must contain a new strongly coupled sector at the TeV scale, but a recent breakthrough has allowed me to make a lot of qualitative predictions for the spectrum of standard model super-partners.  The models, which are called Pyramid Schemes\cite{pyramid}, have a mechanism that produces large mixing between gauginos and composite adjoint chiral superfields.  As a consequence, they predict heavy gauginos, squarks and right handed sleptons that should be in reach of the LHC, and a very complicated Higgs sector, whose properties are hard to extract from the brown muck of the new strong interactions.
One can certainly get a $125$ GeV Higgs, but it is not clear that its interactions are close enough to the standard model to fit the data.  Questions of whether the weak scale is fine tuned in these models are beset by similar strong interaction obscurities.   

The Pyramid Schemes also provide a novel solution to the strong CP problem, a novel dark matter candidate, a possible connection between the dark matter density and baryogenesis, and a possible pathway to explaining extra dark radiation (if the data indicating dark radiation improve to the point where it {\it needs} explanation) .  They explain the absence of all dimension $4$ and $5$ operators which could mediate unobserved baryon and lepton violation, while permitting the dimension $5$ seesaw operator, which gives rise to neutrino masses.

\section{Cosmological SUSY Breaking}

The zero energy generators of the super-BMS algebra provide a huge set of degrees of freedom, localized on the horizon (the conformal boundary ) of Minkowski space that are not incorporated in quantum field theory.  They decouple from the S matrix of particles in Minkowski space. The basic idea of CSB is that the coupling between these states and particles remains finite, in the finite causal diamond of a single geodesic in de Sitter (dS) space.  We view the radius of dS space as a tunable parameter\footnote{The question of what determines this radius in the real world goes beyond the bounds of this review, but has been addressed in \cite{holocosm}.  Briefly, that cosmological model provides a multi-verse of possibilities for the dS radius, and the choice is made by invoking the anthropic principle.} and ask how the coupling between particles and the horizon DOF leads to SUSY breaking.  

To proceed we will have to understand a bit more about the geometry of dS space.  The most important fact about dS space is that even a hypothetical observer, who lives for an infinite amount of time, can only see a finite distance $R$ away.  The entire history of the universe takes place inside a sphere whose radius can never be bigger than $R$.  That sphere is actually running away from the observer at the speed of light.  What is peculiar about the dS universe is that the expanding sphere describing where the backward light-cone from time $T$ meets the forward light-cone from time $- T$, has a finite radius, even as $T$ goes to infinity.  This remains true if the universe is not exactly dS space, but began a finite time in the past, and becomes dS space in the asymptotic future. If observations on the acceleration of the universe are given their simplest interpretation in terms of a positive value for Einstein's cosmological constant (c.c.), then this is what is going on in the universe we live in.  The radius of our {\it cosmological horizon} is about $10^{61}$ Planck units\footnote{The Planck distance scale is $L_P = 10^{-33}$ cm. .  In units where $\hbar = c = 1$ that corresponds to about $T_P \sim 10^{-44} $ seconds, and $M_P = 10^{19}$ GeV.  In these units, Einstein's c.c. is about $10^{ - 123} M_P^4$.}.

Now we want to think about implementing the idea that SUSY breaking comes from interactions with the cosmological horizon in effective field theory (EFT), the framework into which all physics below the Planck energy scale has been assumed to fit.  In effective field theory, SUSY is a gauge symmetry and can only be broken spontaneously.  Gauge symmetries have space-time dependent parameters $\epsilon (x)$.  We can do a gauge transformation on {\it any} Lagrangian.  If the original Lagrangian was not gauge invariant, the result is a new Lagrangian, where $\epsilon (x)$ is a new field.   This new Lagrangian {\it is} gauge invariant, if we let both the original fields and $\epsilon (x)$ transform under the gauge transformation\footnote{This is straightforward for a gauge group with one parameter.  If there are multiple parameter $\epsilon_a (x)$ and the different transformations do not commute with each other the formalism is more complicated, but the results are the same\cite{tbqft} .  }.   The gauge potential is a field $A_{\mu}^a (x)$.  If the original Lagrangian was not gauge invariant, the semi-classical expansion reveals a {\it massive} excitation of the gauge field, and we say that the Schwinger-Anderson-Higgs-Brout-Englert-Guralnik-Hagen-Kibble phenomenon has taken place. In the case of supersymmetry, the gauge parameters are a fermionic spinor field $\epsilon_{\alpha} (x)$.  The massive excitation is that of the gravitino field $\psi_{\mu\alpha} (x)$ and has spin 3/2.  It's mass is denoted $m_{3/2}$.

The mass of the gauge excitation is proportional to the gauge coupling.  In the limit that the gauge coupling goes to zero, the Higgs phenomenon morphs into the phenomenon of Nambu-Goldstone spontaneous symmetry breaking (whence the somewhat inaccurate name spontaneous breaking of gauge symmetry for the Higgs phenomenon). 
The longitudinal part of the massive gauge field becomes the Nambu-Goldstone particle associated with symmetry breaking. The mass is the product of the gauge coupling, and the value of an order parameter, $F$, whose size depends on the energy scale at which the symmetry is broken in the limit of zero gauge coupling.  In the case of supersymmetry, the gravitino is the symmetry partner of the graviton, and the relevant coupling is $\frac{1}{m_P}$\footnote{$m_P = \frac{M_P}{\sqrt{8\pi}}\sim 2\times 10^{18}$ GeV, is called the {\it reduced Planck mass.}.}  The order parameter $F$ has dimensions of squared mass, and determines the typical difference in squared masses between  bosons and their supersymmetric partner fermions, in non-gravitational supermultiplets. The gravitino mass is $$m_{3/2} \sim \frac{F}{m_P} .$$

Super-Poincare invariance appears naturally in $N = 1$ SUGRA only in the presence of a gauged discrete complex R symmetry, which sets the constant in the super-potential to zero.  Indeed, super-symmetry is compatible with general negative values of the cosmological constant.  This has been abundantly confirmed by the AdS/CFT correspondence, in which the quantum theory of many space-times of the form
$$AdS_d \times {\cal K},$$ where ${\cal K}$ is a compact manifold, and Anti-deSitter space is the maximally symmetric space with negative c.c..  SUSY is incompatible with positive c.c. .  The formula for the c.c. in SUGRA is
$$\Lambda = e^K (|F|^2 - \frac{3}{m_P^2} |W|^2 ) .$$ $F$ is the SUSY breaking order parameter, and $W$ is the order parameter for R symmetry.  R symmetry is the subgroup of the discrete symmetry group of the model, which acts on the generators of the supersymmetry algebra\footnote{More precisely: R symmetry is the coset of the discrete symmetry group $G$ by the subgroup $H$, which leaves the SUSY generators invariant. It can be shown that $H$ is a normal subgroup of $G$, so that $G/H$ is a group, called the discrete R symmetry.}.

CSB depends on the hypothesis that the interactions with the horizon generate R violating terms in the effective action, which in turn lead to spontaneous SUSY breaking.  It is also assumed that the gravitino is the lightest particle carrying R charge.

Then the leading order diagrams that could lead to R violating interactions coming from the horizon, have a gravitino line going out to the horizon and another coming back, violating R symmetry by two units .   These diagrams are all proportional to
$$ e^{ - 2 m_{3/2} R_{dS}} C .$$  We can think of this as a term in second order perturbation theory in the interaction via which the horizon emits and absorbs gravitinos, so that
$$ C = \sum_s  \langle out | V^{\dagger} \frac{1}{E - H} | s \rangle \langle s | V | in \rangle,$$ where the sum is over all states of the horizon with which the gravitino can interact.  The energy denominators are all of order $\frac{1}{R_{dS}}$ because the
$e^{\pi (R_{dS} M_P)^2} $ states of the horizon, live in a band of this size.  

The horizon is a null surface and the gravitino can propagate on it for a proper time of order $\frac{1}{m_{3/2}}$.  According to conventional Feynman diagrams, written in the Fock-Schwinger proper time parametrization, it performs a random walk on the surface, with a step size in proper time given by the UV cutoff.   If, as we will assume, the theory has extra dimensions large in $10$ or $11$ dimensional Planck units, then the step size is given by the higher dimensional Planck scale.  Following Witten\cite{wittenstrong} we will assume that this is the scale of coupling unification which is $2 \times 10^{16}$ GeV.   The entropy of the horizon states with which the gravitino interacts is proportional to the area in $4$ dimensional Planck units that it covers in proper time 
$\frac{1}{m_{3/2}}$, and is $c \frac{1}{M_U m_{3/2}}$ .  Thus the full amplitude is proportional to
$$ e^{- 2m_{3/2} R} e^{ c \frac{M_P^2}{M_U m_{3/2}}}$$

Assuming that $m_{3/2}$ goes to zero as a power of $R_{dS}^{-1} $, we find a contradiction unless the power is precisely $R^{- \frac{1}{2}}$.  Indeed, if the mass is assumed to go to zero more rapidly than this, the formula for R breaking interactions blows up exponentially as $R_{dS} \rightarrow \infty$, while if it goes to zero more slowly the strength of these interactions vanishes exponentially.

Using the reduced Planck mass $M_P^2 = 8\pi m_P^2 $ and the relation
$m_P R_{dS}^{-1} = \sqrt{\Lambda / 3} $, we get
$$m_{3/2}^2 = \frac{4\pi c}{\sqrt{3}} \frac{m_P}{ M_U}\sqrt{\Lambda} .$$
For a unification scale $M_U = 2 \times 10^{16}$ GeV, this gives
$$m_{3/2} = K 10^{-2} {\rm eV} , $$ with $K$ a constant of order one.  The SUSY breaking order parameter is thus
$$F = 2 K \times 10^7 ({\rm GeV})^2 .$$  This is a remarkably low value for the SUSY breaking scale, a fact which drives much of the analysis below.

The structure of diagrams contributing to the R violating terms in the Lagrangian implies that these terms {\it do not} satisfy the usual constraints of technical naturalness, familiar from QFT and perturbative string theory.  Any diagram with more than two gravitinos, or with heavier R charged particle states, mediating between the local vertex and the horizon, will be exponentially suppressed.   Diagrams involving R neutral exchanges with the horizon give contributions which have a finite limit as $R_{dS}\rightarrow\infty$, and are small corrections to terms already incorporated in the $\Lambda = 0$ effective Lagrangian.  

As a consequence, apart from the R symmetry itself, symmetries, or approximate symmetries of the $\Lambda = 0$ model, are also preserved by the R-violating terms.  We exploit this in the following way: we choose the R symmetry to forbid all terms of dimension $4$ or $5$ in the $\Lambda = 0$ model, which violate $B$ and $L$, apart from the dimension $5$ superpotential $$ W_{\nu} = \frac{b}{M_{seesaw}} (H_u L )^2 ,$$ which generates neutrino masses of roughly the right order of magnitude\footnote{We do not attempt to explain why $M_{seesaw}$ is an order of magnitude or so less than $M_U$.  This is a high energy problem.}. Insertions of higher dimension $B$ and $L$ violating operators into a diagram with a pair of gravitino lines going out the horizon cannot generate the lower dimension operators, because the extra gravitino loop is cut-off at the SUSY breaking scale or below, by its space-time structure.  We will see later that, with one extra mild assumption, this mechanism for R violation also provides a novel solution to the strong CP problem.

\section{The Pyramid Schemes}

We now want to build an effective field theory, compatible with current experiment, and with the mechanism of CSB.
It must contain the MSSM, as well as an uncontrained goldstino superfield X\footnote{Unconstrained, because the scale of SUSY breaking can be dialed by varying $\Lambda$, so the dynamics that determines the SUSY breaking scale in the effective Lagrangian, must be visible at low energy.}, and must preserve SUSY and a discrete R symmetry, but spontaneously violate SUSY when
R breaking terms are added.   It must be consistent with the bounds on super-partner masses.

If the only dynamically generated scale in the theory is the QCD scale, it is impossible to do this.   The most general renormalizable Lagrangian is that of the Minimal Supersymmetric Standard Model (MSSM), with the additional superpotential\footnote{The value of the superpotential,$W_0$, at the minimum is the order parameter for spontaneous R symmetry breaking.  In the $\Lambda = 0$ limit, $W_0 = 0$. }.

 $$W_X = g_X X H_u H_d + C(X) ,$$ where $C$ is a cubic polynomial.  This always has supersymmetric solutions\footnote{A meta-stable SUSY violating solution can be acceptable only if the probability for transitions into a Big Crunch by tunneling into the basin of attraction of the SUSic minimum is the inverse recurrence time for dS space\cite{abj}\cite{tbjf}.  There are not enough parameters in the model to engineer this.}.   Even if we could generate a non-zero $F_X$ the gluino mass generated by this model would be too small to be compatible with experimental bounds.
Non-renormalizable corrections to this Lagrangian would be suppressed by powers of $M_U$ or $m_P$ and cannot help with these problems.

To remedy the gluino mass problem, we must include a strongly coupled hidden sector, some of whose fields carry color, in order to generate a coupling between $X$ and the QCD field strength $W_{\alpha}^a$, which can give a large enough gluino mass.  This is the only way to generate a new low energy scale in a natural manner. The scale $\Lambda_3$ of this new strongly coupled sector has to be close to the SUSY breaking scale, since $\frac{F_X}{\Lambda_3}$ will be the natural scale that enters into the formula for the gluino mass, and $F_X$ is so low.
The Pyramid Scheme models we propose, have a natural explanation for this coincidence of scales.

The necessity for new colored particles is potentially problematic, if we wish to preserve coupling unification.   The obvious solution to this is to include complete multiplets of some unified gauge group, but one must also be sure that the gauge couplings have no Landau poles below the unification scale\footnote{Some authors like to preserve coupling unification to two loop accuracy.  Two loop corrections are of the same size as one loop threshold corrections at the unification scale, so I have never been very impressed by the ``better fit" given by the two loop results.  I will try only to preserve the one loop results and the fact that the two loop corrections are small.}.   This puts restrictions on the size of the new strong gauge group.

Seiberg's general analysis of the IR behavior of asymptotically free SUSY gauge theories\cite{nati} enables us to rule out many possibilities.  Initially, I was led to the $N_F = N_C = 5$ theory as the unique possibility that could preserve $SU(5)$ unification, but a careful analysis of two loop effects\cite{jj} showed that the model had Landau poles below the unification scale.  I cannot claim to have made an exhaustive survey, but at the moment the only class of models that survives all of these simple tests are the Pyramid Schemes\cite{pyramid}.   

The Pyramid Schemes utilize Glashow's Trinification\cite{trinification}, an $SU(3)^3 \ltimes Z_3$ subgroup of $E_6$, with $3$ generations of chiral fields in the $(1,3,\bar{3}) \oplus (\bar{3}, 1, 3) \oplus 
(3,\bar{3},1)$ representation. The $Z_3$ permutes the three $SU(3)$ groups and ensures equality of couplings in the symmetry limit. I will have to assume the reader is familiar with this, and refer to the {\it i}th subgroup as $SU_i (3)$.  Color is embedded in $SU_3 (3)$ and the weak $SU(2)$ in $SU_2 (3)$.  $10$ chiral fields of each generation are assumed to obtain mass at the unification scale.  The Higgs fields $H_{u,d}$ also belong to an incomplete multiplet, but we do not specify what it is.  More generally, we make no attempt to explain details of physics at the unification scale.

The quiver diagram of Trinification is a chiral triangle.  The simplest extension of it answering our needs extends the triangle to a Pyramid with triangular base (a tetrahedron). 
The fields connecting the apex of the Pyramid to the base
transform in the vector like representation $T_i \oplus \bar{T}_i \in (F,\bar{3}_i) \oplus (\bar{F}, 3_i ) $ and are called {\it trianons}.  $F$ is some representation of the {\it Pyramid Group} $G_P$.   Both the group and the representation must be fairly small, to preserve standard model coupling unification.

While I don't pretend to have made an exhaustive search, the only examples I've found that work are $G_P = SU_P(k)$ with $k = 3,4$ and $F$ the fundamental representation.  The case $k = 3$ is more attractive in a number of ways.  The minimal R symmetry group that works for $k = 3$ is $Z_8$, compared to $Z_{14}$ for $k = 4$.  Furthermore, there's a natural explanation for the coincidence between $\Lambda_3$ and the scale of SUSY breaking for $k = 3$, and an interesting dark matter candidate.  So far, there are no analogous advantages for $k=4$.  

\subsection{The Singlet Sector}

The R symmetry is chosen\cite{pyramid} to forbid all relevant super-potential terms, which would otherwise be expected to be of order $M_U$ or greater.  Another way to say this is that we insist that the $\Lambda = 0$ theory be technically natural.  The super-potential is thus
$$ W_3 = \sum \kappa_i T_i^3  + \tilde{\kappa}_i \tilde(T)_i^3  + W_{std},$$ where the second term is the familiar standard model super-potential, with no $\mu$ term, and no terms violating $B$ or $L$.  The expression $T_i^3$ refers to the cubic invariant of the $(3_P , \bar{3}_i )$ representation.   The unified group would set all these couplings equal, but unification scale symmetry breaking could easily change that, without ruining the success of one loop gauge coupling unification.

The R breaking diagrams with gravitinos propagating out to the horizon can induce terms of the form
$$W_{1,\slash{R}} = m_i T_i \tilde{T}_i + \mu H_u H_d + W_0.$$   The model has a SUSY preserving minimum both with and without the extra terms, and so does not satisfy the requirements of CSB.  This indicates the necessity  of introducing other low energy fields.  
 
 The simplest way to do this, and perhaps the only one, which doesn't disturb the running of the standard model couplings, is to add singlets under the full gauge group.
 It seems that the minimal number is $3$ fields $S^i$.  There is no reason to assume that the index here transforms under the $Z_3$ of the trinification group, though it is suggestive of interesting structure at the unification scale.   The R symmetry action on the $S_i$ can be chosen so that the trilinear couplings 
 $$ W_S = \frac{1}{6}C_{ijk} S^i S^j S^k + \alpha_i^j S^i T_j \tilde{T}_j + \beta_i S^i H_u H_d ,$$ are allowed.   However, we will also impose an additional discrete symmetry, which does not act on the supersymmetry generators, and ensures that the matrix $C_{ij} = C_{ijk}S^k$ has a zero eigenvalue for any choice of $S^k$. The full $\Lambda = 0$ super-potential, $ W_S + W_3 + W_{std}$ has a SUSic, R symmetric minimum when all fields vanish.
 Although the cubic super-potential has flat directions, these are all lifted by non-renormalizable R symmetric corrections to the Kahler and super-potentials, scaled by the unification or Planck scales.  Finally, with the given field content, the gauge couplings remain small at low energy so that the model preserves both SUSY and R symmetry. Thus, this low energy model is consistent with being the low energy limit of the $\Lambda\rightarrow 0$ limit of a model of stable dS space.
 
 We now add 
 
 $$W_{\slash{R}} = m_i T_i \tilde{T}_i + \mu H_u H_d  + \frac{1}{2} m_{ij} S^i S^j + F_i S^i  + W_0,$$  to the superpotential.  The equations for a supersymmetric point become
 
 $$ (\mu + \beta_i S^i) H_{u\ (d)} = 0.$$
 $$\sum_j \alpha_i^j T_j \tilde{T}_j +\frac{1}{2} C_{ijk} S^j S^k + \sum_j m_{ij} S^j + F_i + \beta_i H_u H_d= 0.$$
 $$ (\sum_i S^i \alpha_i^j + m_j) T_j + \tilde{g}_j (\tilde{T})^2_j = 0 .$$
  $$ (\sum_i S^i \alpha_i^j + m_j) \tilde{T}_j + g_j (T)^2_j = 0 .$$
$(\tilde{T})^2_j$ is the bilinear obtained by differentiating the trilinear invariant w.r.t. $\tilde{T}_j$. 

As noted, we can choose the R symmetry, plus another discrete symmetry which does not act on the supersymmetry generators, to ensure that the matrix $$C_{ij} = C_{ijk}S^k ,$$ is not invertible for any $S^i$.  We further assume that the coefficients in $W_{\slash{R}}$ are chosen so that $\mu_{ij}$ shares the zero modes of $C_{ij}$ and that the $S^i$ independent terms in $\frac{\partial W}{\partial S^i}$ have components in the zero mode subspace.  In this case, there can be no SUSic minimum.  The constraints on $W_{\slash{R}}$, do not follow from symmetries, but these terms arise from a very special class of diagrams.  It's only by imposing these constraints that we obtain a low energy model compatible with an underlying gravitational model that breaks SUSY.

To understand fully the dynamics of SUSY breaking in this model, we must first make sure that the $SU_P (3)$ gauge theory indeed becomes strongly coupled, and find the relation between its dynamical scale $\Lambda_3$ and the CSB SUSY breaking scale.   The $SU(3)_P$ Lagrangian at high energies is SUSY QCD, with $N_F = 3 N_C$.  Its one loop beta function vanishes, but in the absence of other couplings the two loop beta function is IR free.  However, if the couplings $g_i$ and $\tilde{g}_i$ are all equal to $\sqrt{4/3} \times$ the gauge coupling, then we have a line of fixed points.  This line is attractive.  We imagine that, at the unification scale $M_U$ the effective theory lies in the domain of attraction of this line and is rapidly sent to a point where the coupling is relatively strong, but barely in the perturbative regime\footnote{The last restriction is imposed in order to be able to do the analysis.  It is likely that even stronger couplings will also work, but it is hard to calculate in that regime.}.  The couplings then remain fixed until we reach the highest mass threshold of the trianon fields. That mass scale is set by the parameters $m_i$, which come from interactions with the horizon.  These three parameters are of comparable order of magnitude and all vanish like $\Lambda^{1/8} m_P^{1/2}$, when the c.c. is sent to zero. We do not know how to calculate them more accurately than that. For phenomenological reasons, we will assume that the lightest mass is $m_3$, the mass of the colored trianon.  

Below the first two trianon thresholds, the lagrangian has $N_F = N_C$ and is asymptotically free.  We have assumed that the fixed line value of the gauge coupling is fairly large, so the confinement scale $\Lambda_3$ is slightly below the masses $m_{1,2}$ of the two colorless trianons.  We can think of the relations between these scales as roughly analogous to that between the charmed quark mass and the QCD scale, $m_i \approx 4\pi \Lambda_3$.  We will also assume that $m_3$ is of order $\Lambda_3$, somewhat analogous to the strange quark mass in QCD.  This means that we can analyze the low energy dynamics in terms of chiral perturbation theory, which in this case means Seiberg's effective Lagrangian\cite{nati}.

The colored trianons are confined into a three by three matrix ${\bf M}$ of {\it pyrmesons}, which transform as an octet and singlet of color, and singlet {\it pyrmabaryon}, $B$, and {\it anti-pyrmabaryon}, $\tilde{B}$, fields.   The effective super-potential on the moduli space is
$$W_{mod} = \Lambda_3^3 [ L (B\tilde{B} - 1 - {\rm det}\ {\bf M} ) + (m_3 + \alpha^3_j S^j) {\rm tr}\ {\bf M} $$
$$+ \kappa_3 B _ {\tilde{\kappa_3}} \tilde{B} + (\beta_j S^j + \mu) H_u H_d + C (S)] + W_{std} + W_0. $$ 
$W_0$ is a constant added in order to tune the c.c. to its observed value. It does not affect the low energy dynamics, which is independent of the Planck mass to first approximation, once we fix the relevant couplings $m_i, \mu,$ and $F_i$.  
In this formula we've rescaled all fields and parameters by powers of $\Lambda_3$, to make them dimensionless. Apart from this rescaling, $C(S)$ is the cubic polynomial in the singlets, that appeared in the super-potential above the scale $\Lambda_3$. $L$ is a Lagrange multiplier field.

Before analyzing the predictions of this model, we note that something very similar results if we set $kappa_i = \tilde{kappa_i} = 0$ for $i = 1$ or $i = 2$.  The UV model no longer has a fixed line, but the couplings vary slowly.  In particular, although the gauge coupling is now IR free, we can still have a strong coupling scale $\Lambda_3$ without producing a Landau pole below $M_U$, as long as $\Lambda_3 < 2 $ TeV\cite{bfk}\footnote{In \cite{bfk} we were trying to preserve predictions of the original Pyramid Scheme, which had explained various lepton excesses, now considered to be due to pulsars.  We preferred schemes where the conserved pyrma-baryon number was carried by colored trianons. The current scenario will not produce light PNGBs.}. 
This is interesting because a model that preserves one of the  pyrma-baryon symmetries at the renormalizable level, allows the lightest particle carrying this quantum number to be a dark matter candidate if an appropriate asymmetry is generated in the very early universe.  We will discuss this further below.

\section{Crude Estimates of MSSM Super-partner Masses}

The first order of business is to estimate the masses of supersymmetric partners of standard model particles in this model.  Here I've recently been surprised.  I'd initially thought that the Pyramid scheme was a form of direct gauge mediation\cite{dgm}.   In fact, for the gluino at least, the gauge mediated masses are much smaller that a contribution from mixing between the gaugino and the pyrmeson octet.  Write $${\bf M} = M + \lambda_a M^a ,$$ where the fields are dimensionless a $\lambda_a$ represent the Gell-Mann matrices.   These fields, are the pyrmesons. Consider an operator
$$ \int d^4 \theta\ f(M, \bar{M}) D_{\alpha} M^a W^{\alpha}_a  + h.c. $$ 
$$ = g_3 f_{MM^*}\frac{F_M \bar{F}_M}{\Lambda_3^3} \psi_{\alpha}^a \lambda_a^{\alpha} + h.c. + fermions + derivatives.$$  
The function $f$ and its derivatives have a factor of the QCD coupling $g_3$, but no loop factors.  They somewhat analogous to hadron magnetic moments in QCD, with the insertion of one weakly coupled field into an effective Lagrangian for composites.  The QCD fine structure constant $\sim .1$ gives $g_3 \sim 1.4$, so this is nominally of the same size as contributions to the Majorana mass 
$$m_3 \int d^2 \theta\ M_a M^a .$$ (These formulae must be rescaled by the Kahler potential to get physical masses), because $F_M$ , $\Lambda_3$ and $m_3$ are all in the TeV range.  

The result is a pair of octet Majorana fermions, whose masses and mass splitting are all of order $\frac{|F_M|^2 }{\Lambda_3^3}$.  Given the rules of CSB, the numerator is bounded by about $10^{14} ({\rm Gev})^4$.  RG running gives an ${\it upper}$ bound on $\Lambda_3 < 2 \times 10^{3}$ GeV.  The lower bound is harder to determine but is related to the fact that we haven't seen any of the pyrma-hadrons and so is probably about $1$ TeV.  Thus, the lightest mass eigenstate with the quantum numbers of the gluino is between $10 - 100$ TeV.  It is a mixture, with order one mixing angle, of the gluino interaction eigenstate and the composite octet fermion in the $N^a$ super-multiplet.  The conventional gauge mediated Majorana gluino mass is suppressed by a factor $\frac{1}{16 \pi^2}$  relative to these masses\cite{tbgluino}.

The dominant contribution to squark squared masses comes from a {\it one QCD loop}, convergent, diagram, as in super-soft models\cite{supersoft}.  The squark masses are universal and are of order $$m_{\tilde{q}} \sim m_{gluino} \sqrt{\frac{g_3^2}{16 \pi^2} }\sim 900 - 9000 {\rm GeV}.$$  The lower reaches of these estimates mean that squarks but not gluinos will be within the reach of the LHC, while the upper values bode ill for near term experimental detection of these particles\footnote{This model has the potential to generate large $F$ terms $F_{u,d} = (\sum \beta_i S^i + \mu) H_{d,u}$, which could splite the stops from the other squarks, one above and one below the universal squark mass.  The strong dynamics makes it hard to predict the values of these terms.}.

A similar formula is also appropriate for the other gauginos, though here the argument is more complicated.  Basically, using arguments analogous to those invoked when discussing mixing between the photon and strongly interacting vector mesons, one counts factors of $g_i$, and $4\pi$, with everything else determined by dimensional analysis and the scale $\Lambda_3$.  In this kind of bilinear mixing, there are no loop factors.   One calculates the two point function of the gauge field strength $W_{\alpha}^i$ and the derivative of the Trianon composite $ D_{\alpha} M^i$, (here $i$ is the adjoint index of the i-th gauge group in the standard model). Then one argues that when $g_i = 0$ there is a stable adjoint fermion, with a Majorana mass of order $2 m_2$ for the $SU(2)$ triplet, and the lighter of $2 m_1$ and $2 m_2$ for the $U(1)$ adjoint (since both of the colorless trianons have $U(1)$ couplings).
Mixing between the gauginos and these states is a seesaw mechanism\cite{seesaw}, giving Majorana gaugino masses of order 
$$m^{(i)}_{1/2} \sim \frac{g_i^2 |F_Y|^4}{2 \Lambda_3^6 m_i}.$$  It's not completely clear which $F$ terms will give the dominant contribution here. The masses are probably less than a TeV, though there's considerable uncertainty in these estimates.  Slepton masses will be down from this by the square root of a loop factor $$\sqrt{\frac{\alpha_i}{4\pi}}.$$
One of the right handed sleptons is thus the NLSP.  If the bino weighs a TeV this crude estimate gives right handed slepton masses of order $30$ GeV, which is already ruled out.  Indeed, for the decay topology of slepton going to Goldstino, sleptons are ruled out up to about $260$ GeV\cite{CMS}\cite{Atlas} and the next run of the LHC might explore another $100$ GeV in mass\footnote{I'd like to thank Scott Thomas and Patrick Draper for explaining the LHC bounds on right handed sleptons to me.}.  If we took our estimates seriously, this would push the bounds on bino and chargino masses up to about $9$ and $27$ TeV.   However the strong $SU_P (3)$ uncertainties do not warrant such drastic conclusions. I cannot emphasize too strongly how much uncertainty there is in these estimates, but they lead us to expect the discovery of right handed sleptons in the near future.

There is one caveat to the claim that a RH slepton is the NLSP, since we have not studied the masses of all the states in the singlet sector.  However, the diagrams contributing to slepton masses are lepton flavor blind.  The coupling of sleptons to the singlet sector is mediated by the Higgs boson and we know that the Higgs couplings of the leptons are small, ranging from $10^{-5}$ to $10^{-2}$.  Decays of a slepton into a lepton and a hypothetical light singlino, would occur outside the LHC detectors.  Thus, even if it turns out that the NLSP is a singlino, the light right handed sleptons predicted by the model should be observed.

\section{Pyramidal Cosmology}

\subsection{Dark Matter}

In the original paper on the Pyramid Scheme, Fortin and I got caught up in the excitement surrounding positron excesses and other dark matter signatures. The majority opinion seems to be that these excesses are no longer considered to be relevant to dark matter.  Since then I've returned to the simple idea\cite{bmo} that dark matter is one of the pyrma-baryons of the strongly coupled Pyramid sector.  This requires that we omit one pair of trilinear couplings from the underlying Lagrangian, and one can choose the R symmetry properties of the model to make this natural.  We have seen that the attractive RG structure of the model is preserved when we do this, as long as the scale $\Lambda_3 < 2$ TeV.

We've seen above, that we want to keep the trilinear couplings of the colored trianon.  This implies that (if the dark matter is the fermion in the supermultiplet) dark matter has a magnetic moment.  This is an old idea, which goes back to technicolor\cite{technibaryondm} and has potential observational consequences\cite{bftdmt}.

In order to get the right dark matter density, we need to postulate an asymmetry generated in the very early universe. This is very easy to do, but has no predictive power.  However, it opens the door to a connection between the dark matter density and the baryon asymmetry of the universe.  The standard model couplings of the trianons lead to a coupling
$$ \frac{\alpha_2^2 }{\Lambda_3^2} J_{PB}^{\mu} J_B^{\mu} ,$$ which implies that an asymmetry in one quantum number will give rise to a chemical potential for the other.  If this chemical potential is substantial when the interactions that violate the corresponding quantum number go out of equilibrium, then spontaneous (pyrma) baryogenesis will occur\cite{kaplan}\cite{bej}, thus connecting the dark matter and baryon densities of the universe.  

There may also be a possible dark matter candidate in the singlet sector of the model, about which I understand too little to make a definitive statement.  Presumably, if it exists, it would be much more like a WIMP.  If this is the dark matter, we can restore the possibility of UV equality of all the trianon trilinear couplings, which is somewhat more elegant.  

\subsection{Dark Radiation}

The gravitinos in any model implementing CSB are very light and were certainly relativistic at the eras where the CMB and structure formation may indicate the need for more relativistic species.  Standard estimates\cite{pagelsprimack} indicate that such light gravitinos decouple before the electroweak phase transition and contribute much less than a neutrino species to the evolution of the universe.   However, non-thermal repopulation of the gravitinos by late decaying NLSPs, could generate the required excess.  This could only occur if the NLSP was part of the singlet sector, because our bounds on light MSSM super-partners rule out such late decays.  

\section{The Strong CP Problem}

As pointed out in \cite{tbtjt} the Pyramid Scheme provides a novel solution to the strong CP problem.   When $\Lambda = 0$ the model has many $U(1)$ symmetries at the renormalizable level, which allow us to rotate away all CP violating phases except the CKM phase.  This would lead to an axion with a decay constant that has been ruled out be experiment.  However, the R symmetry violating  interactions coming from the horizon break all these symmetries and give the axion a large mass.  Normally, when we try to do this in QUEFT, the $U(1)$ breaking terms re-introduce CP violating phases.  

In the Pyramid scheme, these terms come from a very special class of diagrams, where two gravitinos are exchanged with the horizon.  The part of these diagrams localized near the origin has all the symmetries of the $\Lambda = 0$ theory, and the CP violating $\theta_{QCD}$ induced through the CKM matrix is tiny.   The other end of the gravitino lines is more mysterious, but since it lies on the horizon it is at a very high local temperature, of order the unification scale.  Thus, if the fundamental origin of CP violation is spontaneous breakdown, at scales $\ll M_U$ there will be no CP violation near the horizon.   Thus, the phases in all the R breaking diagrams are small, without either fine tuning or an axion.

\section{The Higgs Potential and the Electroweak Scale}

Neglecting loops involving standard model fields, the Higgs potential in the Pyramid Scheme is

$$K^{i\bar{i}} (\beta_i H_u H_d + \alpha_i^3  M + F_i)(\bar{\beta}_{\bar{i}} \bar{H}_u \bar{H}_d + \bar{\alpha}_{\bar{i}}^3  \bar{M} + \bar{F}_{\bar{i}}) + |\beta_i S^i + \mu |^2 (|H_u|^2 + |H_d|^2) .$$  The Kahler potential depends on the singlet fields through the combinations $\alpha_i^k S^i$, for $k = 1,2$.  This comes from integrating out the colorless trianons.
In \cite{tbtjt} this part of the Kahler potential was calculated in zeroth order perturbation theory in the Pyramid coupling. This approximation is not really justified because the masses of the trianons are just a few times $\Lambda_3$.    The parameters $F_i$ and $\mu$ come from interactions with the horizon.  We expect them to be of order a few TeV, but do not have a way to calculate them with any precision.

One should also include contributions to the Higgs potential from stop loops, and, given the size of $SU(2) \times U(1)$ preserving gaugino masses that we have estimated, loops of TeV scale gauginos.  We will also want to choose the couplings $\beta_i$ to be fairly large, which means that loops of singlets will also be important in determining the Higgs potential.

In \cite{tbtjt} we included some, but not all of these effects, many of which push in opposite directions.  We found that we could fit the LHC bounds but that this required a few percent tuning.  Given our new insights into gaugino masses, and the singlet loops, which we simply forgot in \cite{tbtjt} , the problem becomes more complicated.  In addition, the large $\beta_i$ present us with the possibility of large mixing between singlets and the lightest Higgs.  Neglect of the complicated dependence of the Kahler potential on the $S^i$ was unjustified.

Note that the tuning in the Pyramid scheme is not really the same as the oft discussed little hierarchy problem of the MSSM.  It comes from the fact that the Higgs potential above contains a number of relevant parameters whose natural scale in CSB is multiple TeV.  The dimensionless parameters are bounded from above in order to avoid Landau poles below the unification scale.   On the other hand, we have a rather complicated function of $6$ complex variables (the neutral Higgs fields $h_{u,d}$, the singlet pyrmeson $M$, and the three $S^i$ fields ) to minimize, 
so it seems premature to conclude that a tuning of one part in a hundred is unnatural.   It is, at any rate, too complex to attempt here.

\section{Conclusions}

The Pyramid Schemes are the only low energy effective field theories compatible with both the very low scale of SUSY breaking required by CSB, extant experimental data, and standard model gauge coupling unification.  They contain a new strong coupling gauge theory, with fields carrying standard model quantum numbers.  The most attractive candidate so far has a strongly coupled $SU_P(3)$ gauge group.

The strong interactions complicate the computation of the Higgs potential and parts of the spectrum, but terms that give rise to TeV Dirac masses for gluinos (and probably the electroweak gauginos as well) enable us to make a few robust predictions

\begin{itemize}

\item The MSSM spectrum can be characterized as ``flipped mini-split SUSY", with squarks and sleptons systematically lighter than gauginos.  Gluinos will probably not be detected at the LHC, but squarks should show up in the next run, with production and decay modes characteristic of the gluino decoupling limit.  The entire Higgs sector is complicated by mixing with the singlet fields in the low energy model.  This spectrum of colored super-partners is predicted by the model.  It's realized more generally in any model in which there are adjoint chiral superfields, Dirac masses comparable to the supersymmetric adjoint mass term, and small SUSY breaking Majorana terms for the gaugino.  Models with adjoint fields that are elementary up to scales much larger than the SUSY breaking scale, will have problems with gauge coupling unification.

\item The NLSP is either a right handed slepton, or something from the singlet sector, but in any case the right handed sleptons are ``detector stable" and should be seen soon at the LHC, since they decay to leptons and very light gravitino LSPs.  The crudest calculations put their masses $9$ times lower than the LHC lower bound.  The simplest way to solve this problem is to assume that the bino and charged winos are at $9$ and $27$ TeV, but there is so much uncertainty in these estimates from hidden sector strong interactions that one should not take these drastic values that seriously.  Anyone who has followed my work on the Pentagon and Pyramid schemes will know that I'd previously estimated that the bino was the NLSP and that charginos should be found at the LHC.  The recent discovery of operators that give Dirac gaugino masses has changed everything in a dramatic way.

\end{itemize}
Apart from that, the Pyramid Schemes retain the 
flavor structure of gauge mediation\cite{dgm}.  The only violation of rotation symmetries among the generations comes from Standard Model Yukawa couplings, and the mechanism determining the pattern of those is assumed to operate at very high scale.  Dimension $4$ and $5$ baryon and lepton number violation is eliminated by a combination of the discrete R symmetry of the $\Lambda = 0$ model and the special properties of the R breaking operators coming from the horizon.  A similar conspiracy solves the strong CP problem.
The discrete R symmetry imposes an accidental $U_{PQ} (1)$ Peccei-Quinn symmetry on the renormalizable terms of the $\Lambda = 0 $ theory, and the special nature of discrete R violation, combined with the assumption that CP is spontaneously broken at a scale below the unification scale, guarantee that the would be axion is lifted to a high mass, without introducing new phases into low energy couplings.

The Pyramid Schemes also have interesting implications for cosmology.  If we assume one of the pyrma-baryon symmetries is preserved at the renormalizable level, then the dark matter candidate is a standard model singlet fermion, with a mass of $10$s of TeV and a commensurate magnetic dipole moment.  The correct dark matter density is obtained 
by assuming an appropriate primordially generated asymmetry, and there is a potential connection between the dark matter density and the ordinary baryon asymmetry via a form of spontaneous baryogenesis\cite{bej}.

On the other hand it is possible, though not guaranteed, that there can be a light state in the singlet sector that could serve as dark matter.  In this case we would be able to have an elegant and symmetric theory at the unification scale, which would explain the coincidence of scales between $\Lambda_3$ and SUSY breaking.   The model with only two of the three renormalizable PB violating couplings does the same job, but is less elegant.

If the singlet dark matter candidate were sufficiently light, it could be the NLSP, and its stability only due to R parity. Then  it could also be a form of late decaying dark matter, which would produce a dark radiation density in the form of non-thermal gravitinos.  It may be that cosmological data will eventually require us to explain such a density of dark radiation.  The very light gravitinos of the Pyramid Scheme are hard to detect, but beg to be used as dark radiation.  Much more investigation along these lines is needed.

\vskip.3in
\begin{center}
{\bf Acknowledgments }
\end{center}
 I would like to thank M.Dine, W. Shepherd, P. Draper, S. Thomas, D. Shih, L. Carpenter, J-F Fortin, and H. Haber, for numerous conversations about the topics covered in this review. This work was supported in part by the Department of Energy.

\end{document}